# THROUGHPUT PERFORMANCE OF 2×2 MIMO LTE DOWNLINK IN A SPATIAL CORRELATION BASED MICROCELLULAR CHANNEL FOR WIRELESS BROADBAND NETWORKS

Sunil Joshi[1], Deepak Gupta[2], P.C. Bapna[1], Neha Kothari[1], Rashmi Suthar[1]

[1]Department of Electronics & Communication, CTE, MPUAT, Udaipur, India,
`suniljoshi7@rediffmail.com`

[2]Department of Electronics, V.B. Polytechnic College, Udaipur, INDIA,
`simptech@dataone.in`

## ABSTRACT

*Multiple Input Multiple Output (MIMO) technology is going to be a viable alternative for future generation wireless broadband services in order to meet the striving requirements for throughput and system robustness. In Long Term Evolution (LTE), MIMO technologies have been broadly used to get better downlink peak rate, cell coverage, as well as average cell throughput. In the present paper a 2x2 MIMO is taken as baseline configuration for a LTE downlink under a Microcellular propagation scenario considering a non physical correlation based channel with Poor and rich scattering environment. The throughput capacity of the downlink is obtained for poor and rich scattering environments. Besides, two vital aspects of MIMO technique viz Spatial Multiplexing (SM) and Transmit Diversity (TD) are investigated in order to see their effect on throughput of the system. The effect of parameters like Speed of mobile station, number of Multipath, Rician factor (K) on throughput of such systems is reported and discussed. The investigations reported in this paper helps in estimating the throughput capacity of LTE downlink under SM and TD mode.*

## 1. INTRODUCTION

In wireless communications, spectrum is an important resource and hence imposes a high cost on the high data rate transmission. Fortunately, the emergence of multiple antenna system has opened another very resourceful dimension – space, for information transmission in the air. A number of studies have demonstrated that multiple antenna system provides very promising gain in capacity without increasing the power and spectrum [1]. The 3[rd] Generation Partnership Project (3GPP) recommends specifications of Long Term Evolution (LTE), MIMO technologies have been broadly used to get better downlink peak rate, cell coverage, as well as average cell throughput. The main goals of LTE include improving spectral efficiency in 3G networks, allowing carriers to provide more data and voice services over a given bandwidth, lowering costs, improving services, making use of new spectrum and reframed spectrum opportunities, and better integration with other open standards. LTE has introduced a number of new standards to allow IP based wireless mobile broadband. The hybridization of Multiple input multiple output (MIMO) technologies with LTE has generated features such as spatial multiplexing, transmit diversity, and beam forming for better speed and efficiency to support future broadband data service over wireless links [2,3]. The consequence of mobile speed on the performance of 2X2 LTE Downlink MIMO is established in a study reported by Arne Simonsson [4]. He quantified the performance deprivation with reference to the speed of the mobile. The present paper summarizes the studies undertaken pertaining to 2x2 MIMO for LTE downlink standard. The performance of the down link is analyzed under various multi path fading conditions under microcellular propagation environment. The LTE adopts two major MIMO technologies i.e. Spatial Multiplexing (SM) and Transmit diversity (TD). Spatial multiplexing allows transmitting different streams of data simultaneously on the same downlink resource block(s) this increases the data rate of the user [5]. In Transmit Diversity a single stream of data is assigned to the different layers and coded using Space Time Block Coding (STBC). STBC achieves robustness through temporal diversity by using different subcarriers for the repeated data on each antenna [6]. For the LTE downlink, a 2x2 configuration for MIMO is assumed as baseline configuration, i.e. 2 transmit antennas at the base station and 2 receive antennas at the terminal side. Besides, correlation Matrix, a vital channel parameter, is analyzed for of Non – Physical correlation based wireless channel under Poor and rich scattering environment.

## 2. 2×2 MIMO CORRELATION CHANNEL THEORY

A non-physical correlation channel model [10,17] is considered to describe MIMO Correlation model, generally site-independent and are mostly used for system design, comparison and testing. Correlation-based modeling refers to the spatial correlation present among multipath components arriving with different angle of arrivals and powers at the receiver. The Spatial Correlation matrix can be represented by Kronecker product technique as given by [7]:

$$R_{Spot} = R_{BS} \otimes R_{MS} = \begin{bmatrix} 1 & \alpha \\ \alpha^* & 1 \end{bmatrix} \otimes \begin{bmatrix} 1 & \beta \\ \beta^* & 1 \end{bmatrix} = \begin{bmatrix} 1 & \beta & \alpha & \alpha\beta \\ \beta^* & 1 & \alpha\beta^* & \alpha \\ \alpha^* & \alpha^*\beta & 1 & \beta \\ \alpha^*\beta^* & \alpha^* & \beta^* & 1 \end{bmatrix}$$

Where,

$R_{BS} = \begin{pmatrix} 1 & \alpha \\ \alpha^* & 1 \end{pmatrix}$ for the base station and, $R_{MS} = \begin{pmatrix} 1 & \beta \\ \beta^* & 1 \end{pmatrix}$ for the Mobile Station. The values of α and β can be selected to represent different types of channels, and often real values in the range from 0-1 are used [7].

## 3. SYSTEM DESIGN AND PARAMETERS

The schematic diagram of the system is shown in figure 1. The LTE downlink for 2x2 MIMO with two transmit and two receive antenna is designed using Agilent's SystemVue tools. A typical channel bandwidth of 10 MHZ with carrier frequency of 2 GHz is selected with FDD duplex scheme and OFDMA access scheme for downlink. The modulation type supported is 16 QAM, a superior modulation scheme for 2x2 MIMO applications [8, 11]. The Physical Downlink Shared Channel (PDSCH) is used as downlink for data transportation across the LTE radio interface. We have used a correlation channel with finite discrete multipath components which are considered to be uniformly distributed about the transmitter and receiver. Each multipath component is considered uncorrelated and characterize by angle of arrival, angular spread and path gain. Two multipath scenarios, first rural one with poor scattering and second urban microcellular with rich scattering environment are investigated. The antenna-to-antenna spacing at transmitting end is assumed to be 2 λ and 0.5λ at receiving end.

International Journal of Wireless & Mobile Networks (IJWMN) Vol. 3, No. 5, October 2011

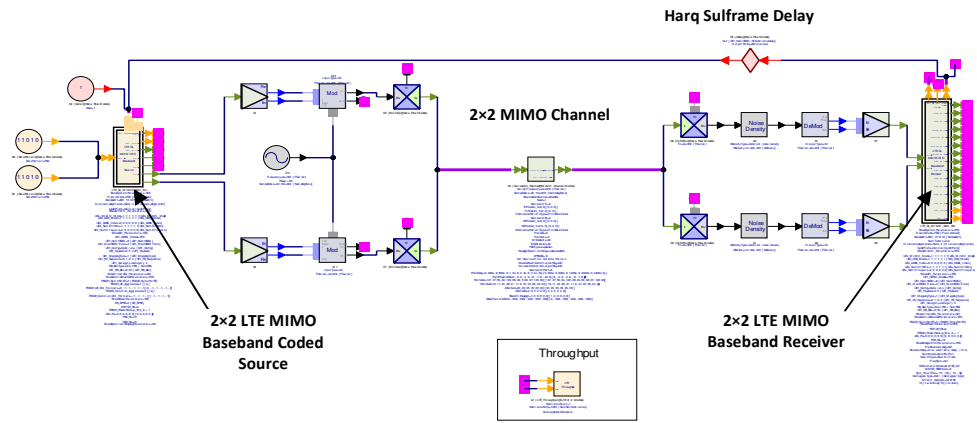

Figure 1. 2×2 MIMO LTE Schematic Design

## 4. RESULTS & DISCUSSIONS

This model performs the averaged closed-loop HARQ throughput over sub frames, from 'Subframe Start' to 'Subframe Stop' for both PDSCH and PUSCH channel [8]. Each firing, one token is consumed at both 'TBS' port and 'CRC Parity' port. The data input from the 'TBS' port indicates the transport block size for each subframe. If the input value at this port is '0', it is assumed that no transport block is allocated in this subframe. The data input from the 'CRCParity' port is the CRC check result for each subframe, where '1' means CRC check is successful and '0' means CRC check fails[9] a.) The throughput fraction of the system designed in presence of poor and rich scattering is investigated. A Poor scattering scenario is assumed to have 4 multipath components representing a rural propagation environment and a Rich scattering scenario is assumed to have 12 multipath components representing an Urban microcellular Propagation environment [17,18]. Besides the scenario Rayleigh distribution and rician distribution by varying the values of K factor from 0 to 6 are also investigated to see their effect on throughput performance of the system.

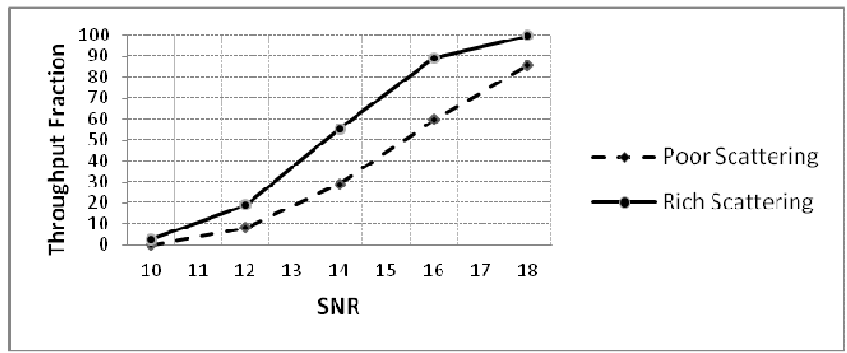

Figure 2.1. Throughput fraction v/s SNR for Spatial Multiplexing ( Rayleigh Distribution )

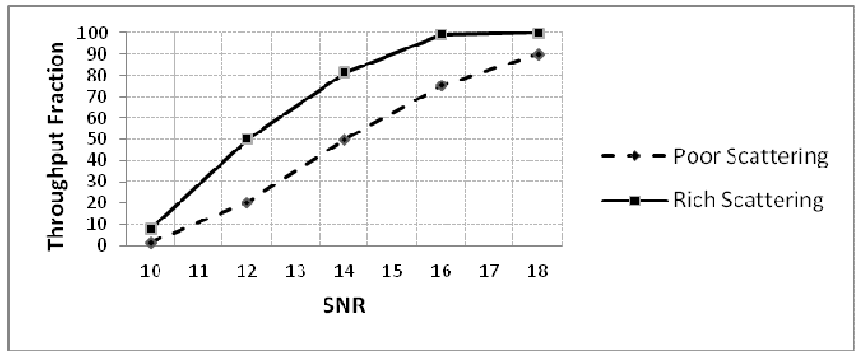

Figure 2.2. Throughput fraction v/s SNR for Spatial Multiplexing (Rician Distribution K=6)

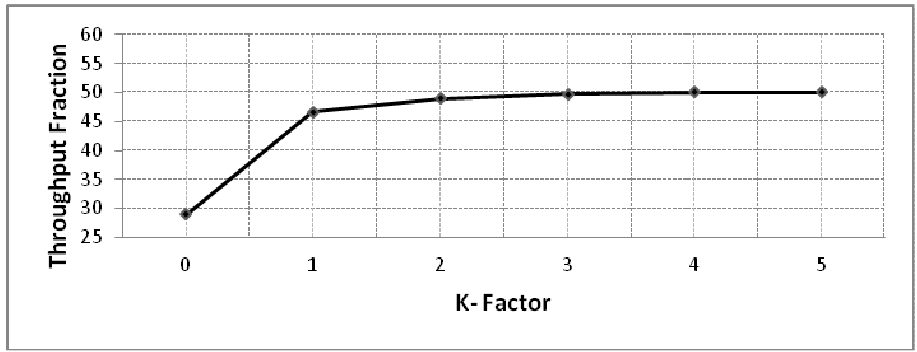

Figure 2.3. Throughput fraction v/s K- Factor for Spatial Multiplexing in poor scattering environment (SNR 14: dB)

The Figure 2.1 reveals that in noisy channel for SNR below 12dB, multipath do not have any significant effect on Throughput fraction. One of the interesting results of SM reveals that the throughput fraction increases almost by two fold when propagation environment is shifted from poor scattering to rich scattering. One of the critical observations derived from fig 2.2 is that under spatial multiplexing mode at relatively low SNR (well below 13dB) the rician factor 'K' is the deciding factor of throughput of the system. Fig 2.3 shows that throughput increases almost linearly when value of K changes from 0 to 1.

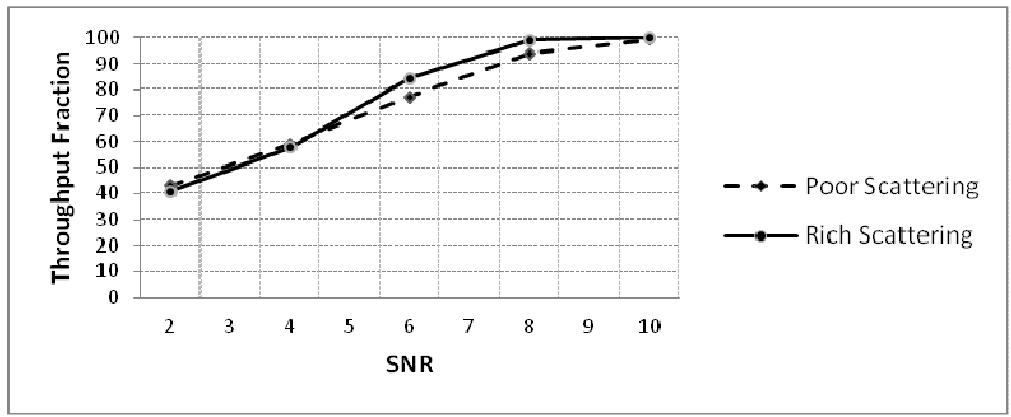

Figure 2.4. SNR v/s Throughput fraction for Transmit Diversity( Rayleigh Distribution )

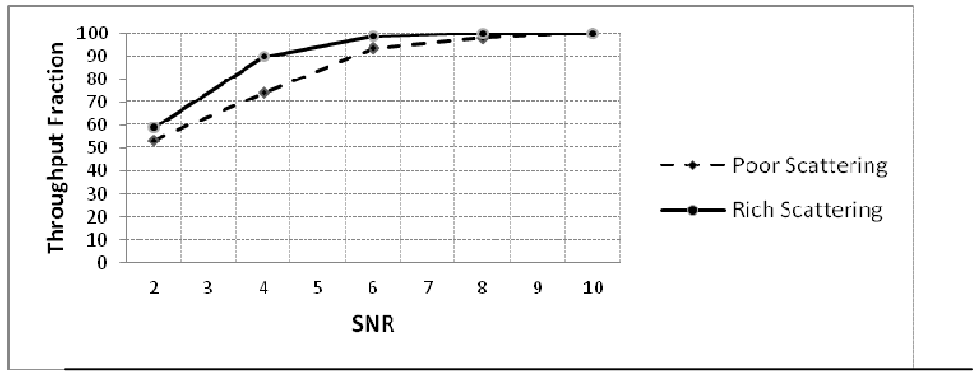

Figure 2.5 SNR v/s Throughput fraction for Transmit Diversity (Rician Distribution K=6)

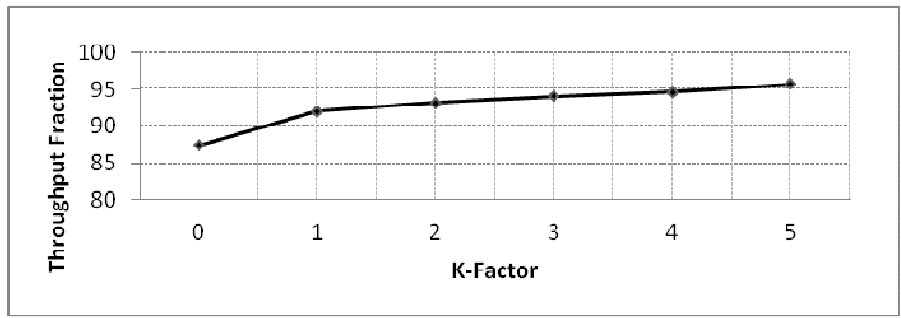

Figure 2.6. Throughput fraction v/s K-factor for Transmit Diversity in poor scattering environment (SNR: 7dB)

Fig 2.4 reveals that in TD mode no. of multipath do not have a significant effect on throughput fraction. As channel conditions improves throughput remains constant (well above 6 dB).Fig 2.5 and 2.6 shows that Presence or absence of LOS component does not affect the throughput fraction.

b.) In mobile environment the vehicle speed is a key factor affecting the data rate of the system. Four vehicle speeds 3, 20, 40, 60Km/hr are separately undertaken to see its effect on throughput performance of the system.

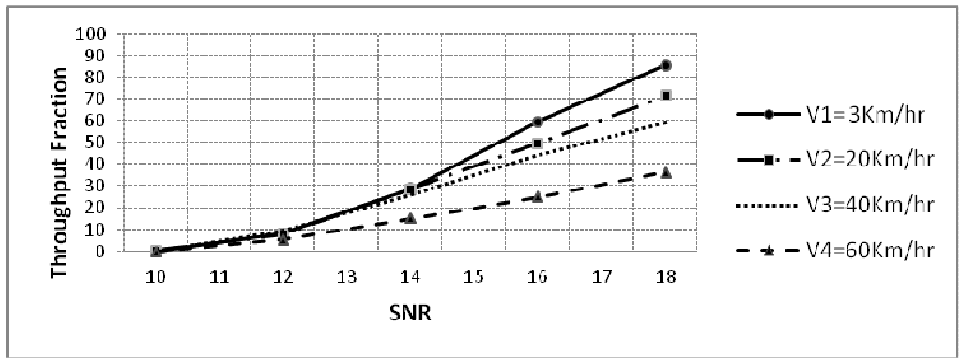

Figure 3.1. SNR v/s Throughput fraction for Spatial Multiplexing with speed variation in Poor scattering environment

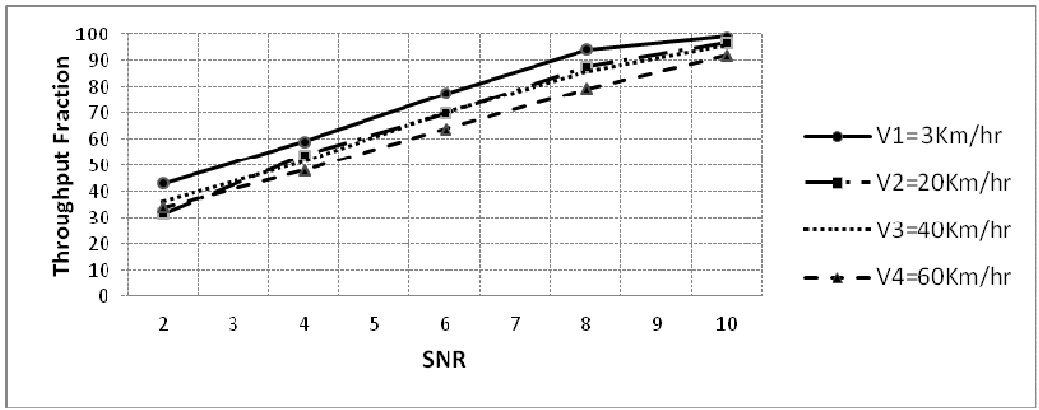

Figure 3.2. SNR v/s Throughput fraction for Transmit Diversity with speed variation in Poor scattering environment

As shown in figure 3.1, in Spatial multiplexing mode, vehicular speed is a key parameter in deciding the throughput of the system. As the speed of Mobile station increases from 20Km/hr to 60Km/hr throughput decreases to about 50%. However, figure 3.2 reveals that in transmit diversity mode mobility of receiver doesn't affect the system throughput.

c) The channel capacity of the system under investigation is sensitive to the multipath environment for different fading distributions. Fig 4.1 and 4.2 indicate channel capacity variations with SNR for Spatial Multiplexing and Transmit Diversity respectively.

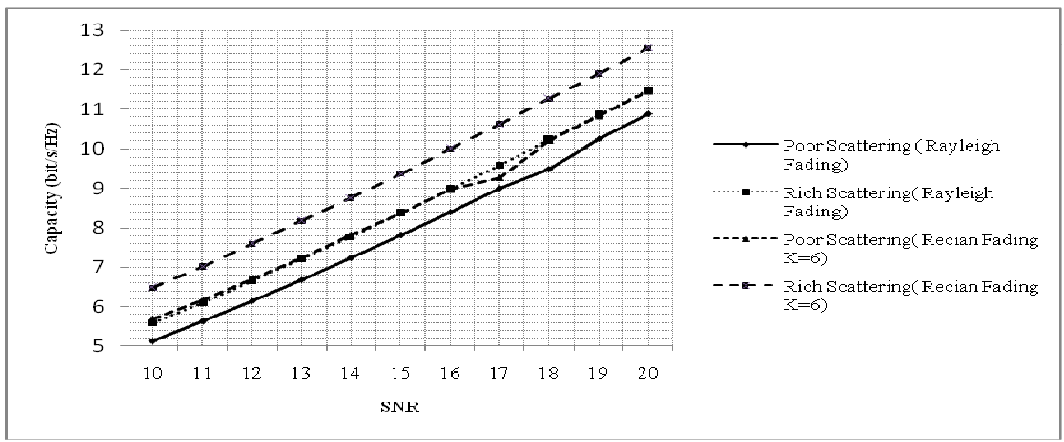

Figure 4.1. Capacity v/s SNR for Spatial Multiplexing

From figure 4.1 it is observed that in spatial multiplexing case the channel capacity obtained is highest in rich scattering environment. A linear variation of capacity with SNR is obtained. In the absence of LOS one can notice that capacity in poor scattering environment is 20% lower than rich scattering for a particular SNR, whereas presence of LOS component enhances the capacity for both the environments up to 17%.

From figure 4.2 it is revealed that in Transmit diversity case also the channel capacity obtained is highest in rich scattering environment. In the absence of LOS one can notice that capacity in poor scattering environment is 14% lower than rich scattering for a particular SNR, whereas presence of LOS component enhances the capacity for both the environments up to 22%.

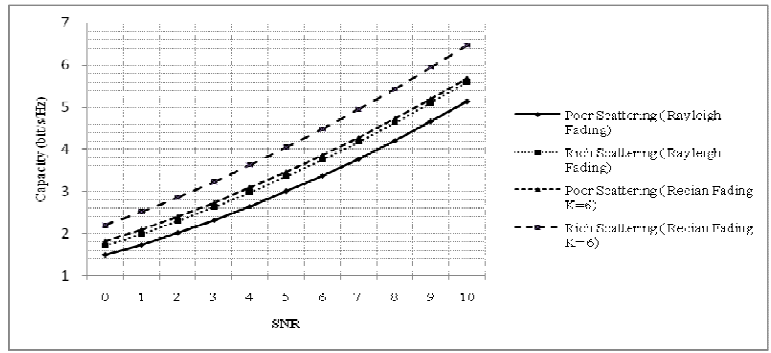

Figure 4.2. Capacity v/s SNR for Transmit Diversity

## 5. CONCLUSION

We have investigated the throughput fraction of 3GPP- Long Term Evolution (LTE) downlink system under two popular modes i.e. Spatial Diversity which is commonly used for quality improvement and Spatial Multiplexing which on the other hand is popular in increasing the data rate. These two investigations were in the presence of poor as well as rich scattering channel conditions. The observation were bold and enough that the Presence of rich scattering environment enhances the system performance. In order to gain full advantage of spatial multiplexing the Signal to Noise Ratio (SNR) should be reasonably good while on the other hand transmit Diversity is a better contender for systems operating at relatively low SNR. Through the simulated results it can be very well proved that the region near the edge of the cell where signal strength is relatively low, it is preferable to use diversity based signal stream transmission strategies. As it only requires one spatial channel which is to be significantly stronger than that of the other. In contrast to this the region when the signal strength is relatively high near the base station, it is beneficial to use Spatial multiplexing strategies. We have also envisaged that the relative strength of the direct and scattered component very well expressed by the Rician factor "K" provides an indication of the link Quality. Through our findings and simulated investigation we may easily say that strategies like spatial multiplexing are undesirable for high-speed users whereas Transmit diversity strategy is attractive for high-speed users.

## 6. ACKNOWLEDGEMENT

We thank department of Information Technology (CC&BT Group), Ministry of Communication and Information Technology, Govt. of India, for there financial support for the research project File No.14(10)/2010-CC&BT.

## Authors

**Dr. Sunil Joshi**   was born in Jodhpur, India in 1967. He received B.E. (Hons) in 1990 in Electronics & Communication Engineering and M.E. in Year 2000 in Digital Communication from MBM Engineering College, JNV University, Jodhpur (India). He taught at various capacities in Diploma and Degree level technical Institutions. He did his Ph.D. in Year 2006 in "Propagation Studies of Millimeter Waves" from Microwave Systems Group Department of Electronics & Communication Engineering Mailviya National Institute of Technology, University of Rajasthan, Jaipur. Presently he is Head of Electronics & Communication Department in the College of Technology & Engineering, Maharana Pratap University of Agriculture & Technology, Udaipur.

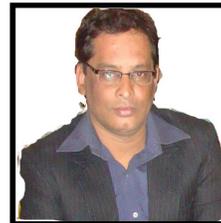

**Deepak Gupta**   was born in Jodhpur, India in 1973. He received B.E. (Hons) in 1995 in Electronics & Communication Engineering and M.E. (Gold Medallist) in Digital Communication from MBM Engineering College, JNV University, Jodhpur (India). He joined as lecturer in Electronics in Vidya Bhawan Polytechnic College, Udaipur in 1996; presently he is Head of Electronics Department in the Institute. He is currently pursing Ph.D. in "Channel Optimization Techniques in Urban Outdoor and Indoor Indian Environment using Multiple Input Multiple-Output Technology".

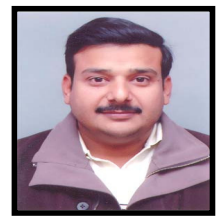

**P.C. Bapna**     was born in Udaipur, India in 1956. He did his engineering studies from University of Engineering, Mysore and Complete his MBA Degree from Mohan Lal Sukhadiya University, Udaipur (Rajasthan). Presently he is Assistant Professor in Electronics & Communication Engineering Department in College of Technology & Engineering, Maharana Pratap University of Agriculture & Technology, Udaipur.  He is currently pursing Ph.D. studies.

**Neha Kothari & Rashmi Suthar**      they did their B.E. in Electronics and communication engineering from  Geetajnali Institute of Technical Studies, Dabok, Udaipur. They are research Scholar in Electronics & Communication Engineering Department in College of Technology & Engineering, Maharana Pratap University of Agriculture & Technology, Udaipur and are working on department of Information Technology (CC&BT Group), Ministry of Communication and Information Technology, Govt. of India project "Channel / Cluster Optimization In Wireless Broadband Access Using MIMO and Multiple Antenna Techniques."

------------------------